\documentclass[preprint2]{aastex6}
\usepackage{graphicx}
\usepackage{natbib}
\usepackage{amsmath}
\usepackage{bm}
\begin{document}

\title{\bf Using Cosmogenic Lithium, Beryllium and Boron to Determine \\ the Surface Ages of Icy Objects in the Outer Solar System}
\author{M. M. Hedman}
\affil{Physics Department, University of Idaho, 875 Perimeter Drive, MS 0903 Moscow ID 83844-0903}

\begin{abstract}
Given current uncertainties in the cratering rates and geological histories of icy objects in the outer solar system, it is worth considering how the ages of icy surfaces could be constrained with measurements from future landed missions. A promising approach would be to determine cosmic-ray exposure ages of surface deposits by measuring the amounts of cosmogenic Lithium, Beryllium and Boron at various depths within a few meters of the surface. Preliminary calculations show that ice that has been exposed to cosmic radiation for one billion years should contain these cosmogenic nuclei at concentrations of a few parts per trillion, so any future experiment that might attempt to perform this sort of measurement will need to meet stringent sensitivity requirements. 
\end{abstract}

\maketitle 

The icy worlds of the outer solar system exhibit a complex array of geological structures that reflect diverse geological histories, but there is still a great deal of uncertainty regarding when various surface features formed. Absolute ages of geological formations on icy bodies have traditionally been estimated based on the observed crater densities for particular surface units. Different models of the impactor flux yield different age estimates, but in general these calculations indicate that the most heavily cratered icy surfaces are comparable in age to the Solar System \citep{Zahnle03, Kirchoff10, DiSisto16, Kirchoff18}. However,  recent studies of the dynamical history of  Saturn's satellites suggest that many of that planet's heavily-cratered mid-sized moons may be substantially less than a billion years old \citep{Cuk16, Asphaug13}. At the moment, it is not clear how to reconcile these dynamical arguments with the moons' cratering records, which highlights how little direct information we have about the ages of solid surfaces in the outer solar system.

Even if future analyses of the currently-available data are able to settle the debates regarding the ages of heavily cratered worlds, there are several objects in the outer solar system that have complex geological histories extending up to the present day, including Europa, Enceladus, Titan, Triton and Pluto \citep{Europa09,  Spencer09, Schenk18, Titan10, Schenk07, Stern15}. For these objects, a key unanswered question is how long they have been active and how long fresh material can be exposed on their surfaces before it is buried or recycled. This question is not only relevant for efforts to understand the geological history of these bodies, it also determines how long ago materials on the surface were in contact with liquid water reservoirs, which has implications for efforts to assess their habitability. It is therefore worth considering what types of future experiments could help constrain the age of icy surfaces. 

The most direct way to measure the absolute age of any solid material is with in-situ measurements of unstable isotopes and/or their daughter products. These sorts of radiometric dates have been obtained from laboratory measurements of both meteorites and lunar samples, and have yielded many important insights into the formation and history of solid bodies in the inner Solar System \citep{Treat01, SR01}. Furthermore, the Mars Science Laboratory recently measured the first radiometric age on another planet, demonstrating that such measurements can be performed by  space missions \citep{Farley14}. There has also been a great deal of recent interest in a mission that would land on Europa and conduct extremely sensitive measurements of surface composition in order to ascertain whether life could exist beneath that world's surface \citep{Hand17}. Hence it is worth examining whether a lander on an icy world could perform experiments  that would yield information about the age of its surface deposits.


The fact that the surfaces of icy bodies are composed primarily of water ice makes many of the commonly-used radiometric dating systems problematic. Even allowing for the possibility that the water ice on various worlds could have substantial amounts of methane, ammonia, and various organic compounds, the elemental composition of their surfaces would still be dominated by the light elements Hydrogen, Carbon, Nitrogen and Oxygen. The longest-lived unstable isotope of these elements is $^{14}$C, which has a half-life of about 5700 years ({\tt http://www.nndc.bnl.gov/nudat2}). While this isotope has been extensively used to date events from the Quaternary period here on Earth ({\tt http://intchron.org}), and has even been proposed as a way to probe carbon transport processes on Titan \citep{Lorenz02}, its half life is far too short to probe the geological history of surfaces that are millions or billions of years old. Indeed, most commonly-used techniques for radiometrically dating rocks involve  nuclei with half lives of order 1 billion years, such as $^{40}$K, $^{87}$Rb and various isotopes of Uranium. While elements with such long-lived unstable isotopes like Potassium could be present near the surfaces of some icy worlds, a dating technique that relies on such contaminants would require more detailed information about the surface composition of these bodies than we currently have. 

In the absence of long-lived isotopes, the most promising way to date ancient icy surfaces is with cosmic-ray exposure ages. The basic idea behind these dates is that any exposed surface is constantly being bombarded with high-energy cosmic rays that cause nuclear reactions within the material. The concentration of the resulting cosmogenic nuclei near the surface therefore increases over time and can provide information about the age of the surface deposit. Such cosmic-ray exposure ages have been used to date various surface deposits on both the  Earth and the Moon, and to determine when meteorites broke free from their parent asteroids  \citep{Eugster03, Dunai10, HC14}.  

Of course, cosmic-ray exposure ages do depend on the assumed flux of cosmic rays, which can vary over time. Indeed, records of cosmogenic nuclei like $^{14}$C and $^{10}$Be in terrestrial ice cores show that the cosmic ray flux here on Earth has varied by roughly a factor of two on timescales of hundreds to thousands of years \citep{Steinhilber12}. Fortunately, these variations appear to average out over the longer timescales relevant for dating astronomical bodies. For example,  analyses of the cosmic-ray exposure ages from stony meteorites (which range between 1 and 100 million years) indicate that the average cosmic ray flux has varied by less than about 10\% over the past 10 million years \citep{Wieler13}.  Exposure ages of iron meteorites extend back 1-2 billion years and can potentially constrain the cosmic ray flux on even longer timescales. However, the interpretation of these data is still uncertain, with some recent analyses suggesting the flux has varied by less than 50\%, while others argue for factor of 3 flux variations over timescales of 500 million years \citep{Ammon09, Wieler13, Alexeev16, Alexeev17}.  Such long-term flux variations would certainly need to be better constrained before cosmic-ray exposure ages could provide accurate geological histories for icy moons. However, even if the long-term  variations were as large as a factor of 3 over 500 million years, cosmic-ray exposure ages could still be used to determine if the heavily-cratered  icy surfaces in the Saturn system are comparable to the age of the solar system or just a few hundred million years old.  

Nuclear reactions induced by high-energy cosmic rays usually generate nuclei with atomic numbers comparable to or less than that of the original nuclei. This again means that only elements with low atomic numbers are likely to be available. Among these, isotopes of Hydrogen, Carbon, Nitrogen and Oxygen are probably not viable because these elements should be common native constituents of the ice, and so distinguishing any cosmogenic material will be extremely challenging. On the other hand, any Helium generated by cosmic rays should diffuse through ice on geological timescales, so  this material  will probably escape the surface. This leaves Lithium, Beryllium and Boron as the most promising elements for cosmic-ray exposure dating of icy surfaces. These elements are all found at very low concentrations in chondrites, Earth's crust and ocean water (see Table~\ref{conc}), and they are all chemically reactive species that can be retained in-situ for geological timescales \citep[Indeed, cosmogenic Beryllium in terrestrial ice cores have been used to trace variations in the cosmic ray flux into Earth's atmosphere over the past 10,000 years, e.g][]{Steinhilber12}.

In order to be able to use cosmogenic Lithium, Beryllium and Boron atoms to constrain the ages of surface deposits on icy bodies, two criteria must be met. First, there must be a way to distinguish atoms produced by cosmic rays from atoms incorporated into the ice by other means. Second, the relevant instrumentation must be sensitive enough to detect the likely concentrations of cosmogenic atoms. As shown below, the first requirement can be achieved by sampling the relevant deposit at several depths within a few meters from the surface, while the latter requirement means that the instrument needs to be able to detect Lithium, Beryllium and/or Boron in water at levels of a few parts per trillion.

\begin{table}
\caption{Number concentration of Lithium, Beryllium and Boron in various materials}
\label{conc}
\hspace{-.5in}\resizebox{3.25in}{!}{\begin{tabular}{|c|c|c|c|} \hline
Material & Lithium & Beryllium & Boron \\ \hline
CI Chondrites$^a$ & $1.5\times10^{-6}$ & $2.5\times10^{-8}$ &  $7\times10^{-7}$ \\
\hline
Bulk Continental Crust$^b$  & $1.6\times10^{-5}$  &  $1.9\times10^{-6}$ & $1.1\times10^{-5}$ \\
\hline
Mean Ocean Water$^c$  & $5\times10^{-7}$  & $4\times10^{-13}$ & $7.5\times10^{-6}$ \\\hline
\end{tabular}}

$^a$ \citet{PJ03}

$^b$ \citet{RG03}

$^c$ {\tt \footnotesize http://www3.mbari.org/chemsensor/summary.html}
\end{table}

Cosmogenic atoms can be identified by measuring concentrations at various depths because the relevant cosmic rays can only penetrate a finite distance through the medium. For example, consider a surface deposit on an icy world that initially has a uniform composition and that does not experience any erosion or overturn. In this situation the concentration of any stable nuclei as a function of time $t$ and depth $z$ should be given by the formula \citep{Dunai10}:

\begin{equation}
C(t,z)= C_i + P_0 t e^{-z\rho/\Lambda} 
\label{conceq}
\end{equation}
where $C_i$ is the initial concentration of relevant atoms, $P_0$ is the production rate of those atoms by cosmic rays at the surface, $\rho$ is  the mass density of the surface deposit, and $\Lambda$ is the effective average ``attenuation length'' of the relevant cosmic rays through that material (note this parameter is actually a column density with units of g/cm$^2$). In this simple situation one can determine the age-dependent cosmogenic concentration factor $P_0 t$ by measuring the concentration over a range  of depths comparable to the scale length $L=\Lambda/\rho$. For a wide variety of materials (including air, ice and rock), the cosmic rays that produce light isotopes like $^{10}$Be and $^{14}$C have a $\Lambda \sim$ 150~g/cm$^2$ \citep{Brown92,Dunai00,Farber08,NN12}, and the density of any sensible surface deposit on an icy body will be between 0.5 and 1 g/cm$^3$. The typical scale length for variations in the amounts of  cosmogenic atoms should therefore be between 1 and 3 meters. 

Of course, real surface deposits on icy worlds will not be completely pristine, and will have instead experienced some amount of regolith overturn due to meteoroid bombardment. This overturn mixes the upper layers of the deposit and so would cause the concentration of cosmogenic nuclei to deviate from the simple form given by Equation~\ref{conceq}.\footnote{This mixing will also carry any ions implanted onto the surface by the solar wind downwards. Note that these atoms only penetrate materials to depths of order a micron \citep{SRIM10}, and so solar wind implantation should not corrupt the cosmic-ray exposure age much below the regolith overturn depth.} Fortunately, studies of the lunar regolith show that the overturn depth is less than half a meter even for deposits that are nearly 4.5 billion years old \citep{Morris78, Blanford80, Costello18}. Hence it is reasonable to expect that regolith overturn on icy worlds will not penetrate so deeply that it will completely erase the vertical trend for cosmogenic atoms. In fact, with sufficiently complete vertical sampling between the surface and a few meters depth it should be possible to ascertain not only the concentration of cosmogenic atoms but also how deeply the regolith is being overturned. 

To obtain a useful age from Lithium, Beryllium or Boron concentrations, one needs a reasonable estimate of the production rate $P_0$, and the instrument conducting this experiment needs sufficient sensitivity to measure concentrations of order $P_0t$.  The production rates for cosmogenic nuclei in different materials have been measured under various conditions on Earth. For the purposes of this report, the most relevant data come from experiments where water tanks were exposed to cosmic rays at high altitudes for a couple of years, after which the amounts of cosmogenic Beryllium isotopes in the tanks were measured to ascertain the appropriate production rates. These experiments indicate that $^{10}$Be is generated in water-rich targets at high-altitudes  at a rate of around 80 atoms per gram of water per year, and $^7$Be atoms  are produced at about half this rate \citep{Nishiizumi96, Brown00}. Note that these specific Beryllium isotopes are unstable and have lifetimes that are much shorter than the likely surface ages of most icy worlds. Specifically, $^{10}$Be transforms into $^{10}$B with a half-life of around 1.39$\times10^6$ years, and $^7$Be decays into $^7$Li with a half-life of about 53 days. These Beryllium isotopes therefore cannot be used directly to date ancient surfaces. Fortunately, the isotopes of Boron and Lithium generated by the decay of these nuclei are stable and so will accumulate in the ice at a rate comparable to the production rate of the Beryllium isotopes over geological timescales.

Of course, the actual production rate of cosmogenic nuclei on the surface of icy bodies in the outer Solar System will not be exactly the same as the production rate in water here on Earth because the cosmic rays responsible for producing these nuclei are partially absorbed by Earth's atmosphere and could be partially deflected by the powerful magnetic fields surrounding the giant planets. Furthermore, the actual production rates for stable nuclei would need to include contributions from all possible pathways (i.e.\ direct formation as well as from decaying unstable isotopes). These factors will therefore need to be evaluated before any precise age estimate could be obtained using this method, but such detailed calculations are beyond the scope of this preliminary work. Fortunately, the relevant production rates probably do not vary by many orders of magnitude across the Solar System. The cosmic rays that can most efficiently produce nuclear reactions are the ones with the highest energy. These particles are the most difficult to deflect or absorb, making their flux less dependent on the specific environment. Hence, for the purposes of obtaining a rough order-of-magnitude estimate of the concentrations of Lithium, Beryllium and Boron near the surfaces of icy worlds we can assume the production rate of each of these elements in ice is comparable to the production rate of unstable Beryllium isotopes in water tanks on Earth.

Assuming that $P_0$ is of order 100 atoms per gram of ice per year and an exposure age of 1~Gyr, the cosmogenic concentration parameter near the surfaces of icy worlds becomes $10^{11}$~atoms (or roughly 0.2 pmol) per gram of ice, which corresponds to a number concentration of 3 parts per trillion, or  200 pmol/kg. An instrument capable of measuring Lithium, Beryllium or Boron concentrations at parts-per-trillion levels should therefore be able to distinguish surfaces formed during the early solar system from deposits formed in the last couple hundred million years.  Such low concentrations have not yet been measured in any spacecraft-based experiment, and indeed the high sensitivity required for this measurement will probably be the single most challenging aspect of this experiment. 

First of all, securely identifying cosmogenic Lithium, Beryllium and Boron based on their decreasing concentration with depth will be difficult if there is a much higher concentration of  native nuclei near the surfaces of icy bodies (i.e. if $C_i>>P_0t$). The average concentrations of Lithium, Beryllium and Boron in Earth's Crust or CI chondrites range from parts per million to parts per billion (see Table~\ref{conc}), which is likely large enough to make identifying cosmogenic nuclei challenging.
 However, the concentration of stable Beryllium in Earth's oceans is below one part per trillion, or around 30 pmol/kg \citep{Measures96, Frank09, Tazoe14}, which is actually less than the expected concentration of cosmogenic Beryllium near a billion-year-old icy surface. At the moment, it is not clear how native Lithium, Beryllium and Boron  would be partitioned within an icy moon, and so additional work is needed to ascertain whether surface ices in the outer solar system are likely to have sufficiently low native concentrations of  these elements.  Nevertheless, Earth's ocean water at least demonstrates that such situations are not an unreasonable possibility. 

Assuming that the native amounts of Lithium, Beryllium and/or Boron are low enough to allow concentrations of cosmogenic nuclei to be identified, the question then becomes whether any landed experiment could achieve high enough sensitivities to reliably measure parts-per-trillion concentrations of these atoms. A detailed assessment of potential methods that could be employed for this experiment is beyond the scope of this report. However, we can note that this dating method only  requires measuring concentrations of chemically distinctive elements like Boron, Beryllium  or Lithium (i.e. isotopic analysis is not required). This means that appropriate chemical separation and concentration techniques could be used to obtain the required sensitivity levels.  Furthermore, the expected concentrations of cosmogenic nuclei in billion-year-old  ice are  comparable to the observed concentrations of stable Beryllium in ocean water \citep{Measures96, Frank09, Tazoe14}.  Methods used to survey the Beryllium in Earth's oceans should therefore provide a useful basis for developing the requisite techniques for this sort of experiment. These terrestrial studies have employed a variety of measurement techniques, including electron-capture detection gas-chromatography \citep{Measures86}, mass spectrometry after pre-concentration in a silica-gel  \citep{Tazoe14} and even using microcantilevers coated with hydrogels \citep{Peng12}. While it is not clear whether any of these methods can be practically adapted to work on a spacecraft mission to the outer solar system, the sensitivities considered here are also not that far beyond some of requirements proposed for a Europa lander. For example, \citet{Hand17} state that a Europa lander should be able to measure ``molecules of potential biological origin (biomolecules or metabolic products) at compound concentrations as low as 1 picomole in a 1 gram sample of Europan surface material".  While the specific chemical techniques needed to isolate organic molecules are very different from those needed to detect cosmogenic nuclei, this at least suggests that such precise measurements could be within the realm of possibility for near-future spacecraft.

In conclusion, while there are some significant challenges to measuring concentrations of cosmogenic Lithium, Beryllium and Boron, this currently appears to be the most promising way to constrain the age of surface deposits on icy worlds. Of course, further work is needed in order to identify experimental techniques that could achieve the desired sensitivities, evaluate the likely amounts of native atoms in relevant deposits, and determine the relevant production rates. Such studies could provide important steps towards understanding the ancient geological history of the outer Solar System.

{\bf Acknowledgements:} The author thanks R.N. Clark, M.S. Tiscareno, G.D. Bart, J.W. Barnes, P.D. Nicholson, L. Baker, S. MacKenzie and R. Chancia for reading through early versions of this manuscript and providing helpful comments. He also thanks R. Lorenz for his helpful review of the draft of this manuscript.


\begin{thebibliography}{}

\bibitem[{Alexeev}(2016){Alexeev}]{Alexeev16}
{Alexeev}, V.~A. 2016.
\newblock {On time variations of the intensity of galactic cosmic rays for the
  recent billion years from the data on exposure ages of iron meteorites}.
\newblock {\em Solar System Research\/}~{\bf 50}, 24--32.

\bibitem[{Alexeev}(2017){Alexeev}]{Alexeev17}
{Alexeev}, V.~A. 2017.
\newblock {A model of possible variations of the galactic cosmic ray intensity
  over the recent billion years}.
\newblock {\em Solar System Research\/}~{\bf 51}, 196--203.

\bibitem[{Ammon} {\em et~al.}(2009){Ammon}, {Masarik}, and {Leya}]{Ammon09}
{Ammon}, K., J.~{Masarik},\ and I.~{Leya} 2009.
\newblock {New model calculations for the production rates of cosmogenic
  nuclides in iron meteorites}.
\newblock {\em Meteoritics and Planetary Science\/}~{\bf 44}, 485--503.

\bibitem[{Asphaug} and {Reufer}(2013){Asphaug} and {Reufer}]{Asphaug13}
{Asphaug}, E.,\ and A.~{Reufer} 2013.
\newblock {Late origin of the Saturn system}.
\newblock {\em Icarus\/}~{\bf 223}, 544--565.

\bibitem[{Blanford}(1980){Blanford}]{Blanford80}
{Blanford}, G.~E. 1980.
\newblock {Cosmic ray production curves below reworking zones}.
\newblock In S.~A. {Bedini} (Ed.), {\em Lunar and Planetary Science Conference
  Proceedings}, Volume~11 of {\em Lunar and Planetary Science Conference
  Proceedings}, pp.\  1357--1368.

\bibitem[{Brown} {\em et~al.}(1992){Brown}, {Brook}, {Raisbeck}, {Yiou}, and
  {Kurz}]{Brown92}
{Brown}, E.~T., E.~J. {Brook}, G.~M. {Raisbeck}, F.~{Yiou},\ and M.~D. {Kurz}
  1992.
\newblock {Effective attenuation lengths of cosmic rays producing Be-10 and
  Al-26 in quartz - Implications for exposure age dating}.
\newblock {\em GRL\/}~{\bf 19}, 369--372.

\bibitem[{Brown} {\em et~al.}(2000){Brown}, {Trull}, {Jean-Baptiste},
  {Raisbeck}, {Bourl{\`e}s}, {Yiou}, and {Marty}]{Brown00}
{Brown}, E.~T., T.~W. {Trull}, P.~{Jean-Baptiste}, G.~{Raisbeck},
  D.~{Bourl{\`e}s}, F.~{Yiou},\ and B.~{Marty} 2000.
\newblock {Determination of cosmogenic production rates of $^{10}$Be, $^{3}$He
  and $^{3}$H in water}.
\newblock {\em Nuclear Instruments and Methods in Physics Research B\/}~{\bf
  172}, 873--883.

\bibitem[{Brown} {\em et~al.}(2010){Brown}, {Lebreton}, and {Waite}]{Titan10}
{Brown}, R.~H., J.-P. {Lebreton},\ and J.~H. {Waite} 2010.
\newblock {\em {Titan from Cassini-Huygens}}.

\bibitem[{Costello} {\em et~al.}(2018){Costello}, {Ghent}, and
  {Lucey}]{Costello18}
{Costello}, E.~S., R.~R. {Ghent},\ and P.~G. {Lucey} 2018.
\newblock {The mixing of lunar regolith: Vital updates to a canonical model}.
\newblock {\em Icarus\/}~{\bf 314}, 327--344.

\bibitem[{{\'C}uk} {\em et~al.}(2016){{\'C}uk}, {Dones}, and
  {Nesvorn{\'y}}]{Cuk16}
{{\'C}uk}, M., L.~{Dones},\ and D.~{Nesvorn{\'y}} 2016.
\newblock {Dynamical Evidence for a Late Formation of Saturn's Moons}.
\newblock {\em \apj\/}~{\bf 820}, 97.

\bibitem[{Davis} {\em et~al.}(2003){Davis}, {Holland}, and {Turekian}]{Treat01}
{Davis}, A.~M., H.~D. {Holland},\ and K.~K. {Turekian} 2003.
\newblock {Treatise on Geochemistry, Volume 1}.
\newblock {\em Treatise on Geochemistry\/}~{\bf 1}, 711.

\bibitem[{Di Sisto} and {Zanardi}(2016){Di Sisto} and {Zanardi}]{DiSisto16}
{Di Sisto}, R.~P.,\ and M.~{Zanardi} 2016.
\newblock {Surface ages of mid-size saturnian satellites}.
\newblock {\em Icarus\/}~{\bf 264}, 90--101.

\bibitem[{Dunai}(2000){Dunai}]{Dunai00}
{Dunai}, T.~J. 2000.
\newblock {Scaling factors for production rates of in situ produced cosmogenic
  nuclides: a critical reevaluation}.
\newblock {\em Earth and Planetary Science Letters\/}~{\bf 176}, 157--169.

\bibitem[{Dunai}(2010){Dunai}]{Dunai10}
{Dunai}, T.~J. 2010.
\newblock {\em {Cosmogenic Nuclides: Principles, Concepts and Applications in
  the Earth Surface Sciences}}.
\newblock Cambridge University Press.

\bibitem[{Eugster}(2003){Eugster}]{Eugster03}
{Eugster}, O. 2003.
\newblock {Cosmic-ray Exposure Ages of Meteorites and Lunar Rocks and Their
  Significance}.
\newblock {\em Chemie der Erde / Geochemistry\/}~{\bf 63}, 3--30.

\bibitem[{Farber} {\em et~al.}(2008){Farber}, {M{\'e}riaux}, and
  {Finkel}]{Farber08}
{Farber}, D.~L., A.-S. {M{\'e}riaux},\ and R.~C. {Finkel} 2008.
\newblock {Attenuation length for fast nucleon production of $^{10}$Be derived
  from near-surface production profiles}.
\newblock {\em Earth and Planetary Science Letters\/}~{\bf 274}, 295--300.

\bibitem[{Farley} {\em et~al.}(2014){Farley}, {Malespin}, {Mahaffy},
  {Grotzinger}, {Vasconcelos}, {Milliken}, {Malin}, {Edgett}, {Pavlov},
  {Hurowitz}, and et~al.]{Farley14}
{Farley}, K.~A., C.~{Malespin}, P.~{Mahaffy}, J.~P. {Grotzinger}, P.~M.
  {Vasconcelos}, R.~E. {Milliken}, M.~{Malin}, K.~S. {Edgett}, A.~A. {Pavlov},
  J.~A. {Hurowitz},\ and et~al. 2014.
\newblock {In Situ Radiometric and Exposure Age Dating of the Martian Surface}.
\newblock {\em Science\/}~{\bf 343}, 1247166.

\bibitem[{Frank} {\em et~al.}(2009){Frank}, {Porcelli}, {Andersson},
  {Baskaran}, {Bj{\"o}rk}, {Kubik}, {Hattendorf}, and {Guenther}]{Frank09}
{Frank}, M., D.~{Porcelli}, P.~{Andersson}, M.~{Baskaran}, G.~{Bj{\"o}rk},
  P.~W. {Kubik}, B.~{Hattendorf},\ and D.~{Guenther} 2009.
\newblock {The dissolved Beryllium isotope composition of the Arctic Ocean}.
\newblock {\em Geochimica et Cosmochimica Acta\/}~{\bf 73}, 6114--6133.

\bibitem[{Hand} {\em et~al.}(2017){Hand}, {Murray}, {Garvin}, {Brickerhoff},
  B.C., {Edgett}, {Ehlmann}, {German}, A.G., T.M., {Horst}, {Lunine},
  {Nealson}, {Paranicas}, {Schmidt}, {Smith}, {Rhoden}, {Templeton}, P.A.,
  {Yingst}, {Phillips}, {Cable}, {Craft}, {Hofmann}, {Nordheim}, {Pappalard},
  and {Project Engineering Team}]{Hand17}
{Hand}, K., A.~{Murray}, J.~{Garvin}, W.~{Brickerhoff}, C.~B.C., K.~{Edgett},
  B.~{Ehlmann}, C.~{German}, H.~A.G., H.~T.M., S.~{Horst}, J.~{Lunine},
  K.~{Nealson}, C.~{Paranicas}, B.~{Schmidt}, D.~{Smith}, M.~{Rhoden}, A.R.
  amd~{Russel}, A.~{Templeton}, W.~P.A., R.~{Yingst}, C.~{Phillips},
  M.~{Cable}, K.~{Craft}, A.~{Hofmann}, T.~{Nordheim}, R.~{Pappalard},\ and
  {Project Engineering Team} 2017.
\newblock {\em {Report of the Europa Lander Science Definition Team}}.

\bibitem[{Herzog} and {Caffee}(2014){Herzog} and {Caffee}]{HC14}
{Herzog}, G.~F.,\ and M.~W. {Caffee} 2014.
\newblock {Cosmic-Ray Exposure Ages of Meteorites}.
\newblock In A.~M. {Davis} (Ed.), {\em Meteorites and Cosmochemical Processes},
  pp.\  419--454.

\bibitem[{Kirchoff} {\em et~al.}(2018){Kirchoff}, {Bierhaus}, {Dones},
  {Robbins}, {Singer}, {Wagner}, and {Zahnle}]{Kirchoff18}
{Kirchoff}, M., E.~{Bierhaus}, L.~{Dones}, S.~{Robbins}, K.~{Singer},
  R.~{Wagner},\ and K.~{Zahnle} 2018.
\newblock {Cratering Histories in the Saturnian System}.
\newblock In P.~{Shenck}, R.~{Clark}, C.~{Howett}, A.~{Verbiscer}, and
  J.~{Hunter Waite} (Eds.), {\em Enceladus and the Icy Moons of Saturn}, pp.\
  267--284. University of Arizona Press.

\bibitem[{Kirchoff} and {Schenk}(2010){Kirchoff} and {Schenk}]{Kirchoff10}
{Kirchoff}, M.~R.,\ and P.~{Schenk} 2010.
\newblock {Impact cratering records of the mid-sized, icy saturnian
  satellites}.
\newblock {\em Icarus\/}~{\bf 206}, 485--497.

\bibitem[{Lorenz} {\em et~al.}(2002){Lorenz}, {Jull}, {Swindle}, and
  {Lunine}]{Lorenz02}
{Lorenz}, R.~D., A.~J.~T. {Jull}, T.~D. {Swindle},\ and J.~I. {Lunine} 2002.
\newblock {Radiocarbon on Titan}.
\newblock {\em Meteoritics and Planetary Science\/}~{\bf 37}, 867--874.

\bibitem[Measures and Edmond(1986)Measures and Edmond]{Measures86}
Measures, C.,\ and J.~Edmond {1986}.
\newblock {Determination of Beryllium in natural waters in real time using
  electron-capture detection gas-chromatography }.
\newblock {\em {Analytical Chemistry}\/}~{\em {58}\/}({9}), {2065--2069}.

\bibitem[Measures {\em et~al.}(1996)Measures, Ku, Luo, Southon, Xu, and
  Kusakabe]{Measures96}
Measures, C., T.~Ku, S.~Luo, J.~Southon, X.~Xu,\ and M.~Kusakabe {1996}.
\newblock {The distribution of Be-10 and Be-9 in the South Atlantic}.
\newblock {\em {DEEP-SEA RESEARCH PART I-OCEANOGRAPHIC RESEARCH PAPERS}\/}~{\em
  {43}\/}({7}), {987--1009}.

\bibitem[{Morris}(1978){Morris}]{Morris78}
{Morris}, R.~V. 1978.
\newblock {In situ reworking /gardening/ of the lunar surface - Evidence from
  the Apollo cores}.
\newblock In {\em Lunar and Planetary Science Conference Proceedings}, Volume~9
  of {\em Lunar and Planetary Science Conference Proceedings}, pp.\
  1801--1811.

\bibitem[{Nesterenok} and {Naidenov}(2012){Nesterenok} and {Naidenov}]{NN12}
{Nesterenok}, A.,\ and V.~{Naidenov} 2012.
\newblock {In situ formation of cosmogenic $^{14}$C by cosmic ray nucleons in
  polar ice}.
\newblock {\em Nuclear Instruments and Methods in Physics Research B\/}~{\bf
  270}, 12--18.

\bibitem[{Nishiizumi} {\em et~al.}(1996){Nishiizumi}, {Finkel}, {Klein}, and
  {Kohl}]{Nishiizumi96}
{Nishiizumi}, K., R.~C. {Finkel}, J.~{Klein},\ and C.~P. {Kohl} 1996.
\newblock {Cosmogenic production of $^{7}$Be and $^{10}$Be in water targets}.
\newblock {\em \jgr\/}~{\bf 101}, 22225--22232.

\bibitem[{Palme} and {Jones}(2003){Palme} and {Jones}]{PJ03}
{Palme}, H.,\ and A.~{Jones} 2003.
\newblock {Solar System Abundances of the Elements}.
\newblock {\em Treatise on Geochemistry\/}~{\bf 1}, 711.

\bibitem[{Pappalardo} {\em et~al.}(2009){Pappalardo}, {McKinnon}, and
  {Khurana}]{Europa09}
{Pappalardo}, R.~T., W.~B. {McKinnon},\ and K.~K. {Khurana} 2009.
\newblock {\em {Europa}}.
\newblock University of Arizona Press, Tucson.

\bibitem[Peng {\em et~al.}(2012)Peng, Chen, Ji, Wu, and Tung]{Peng12}
Peng, R.-P., B.~Chen, H.-F. Ji, L.-Z. Wu,\ and C.-H. Tung {2012}.
\newblock {Highly sensitive and selective detection of beryllium ions using a
  microcantilever modified with benzo-9-crown-3 doped hydrogel}.
\newblock {\em {ANALYST}\/}~{\em {137}\/}({5}), {1220--1224}.

\bibitem[{Rudnick} and {Gao}(2003){Rudnick} and {Gao}]{RG03}
{Rudnick}, R.~L.,\ and S.~{Gao} 2003.
\newblock {Composition of the Continental Crust}.
\newblock {\em Treatise on Geochemistry\/}~{\bf 3}, 659.

\bibitem[Schenk {\em et~al.}(2018)Schenk, Clark, Howett, Verbiscer, and
  Waitr]{Schenk18}
Schenk, P., R.~Clark, C.~Howett, A.~Verbiscer,\ and J.~Waitr 2018.
\newblock {\em {Enceladus and the Icy Moons of Saturn}}.
\newblock University of Arizona Press.

\bibitem[{Schenk} and {Zahnle}(2007){Schenk} and {Zahnle}]{Schenk07}
{Schenk}, P.~M.,\ and K.~{Zahnle} 2007.
\newblock {On the negligible surface age of Triton}.
\newblock {\em Icarus\/}~{\bf 192}, 135--149.

\bibitem[{Spencer} {\em et~al.}(2009){Spencer}, {Barr}, {Esposito},
  {Helfenstein}, {Ingersoll}, {Jaumann}, {McKay}, {Nimmo}, and
  {Waite}]{Spencer09}
{Spencer}, J.~R., A.~C. {Barr}, L.~W. {Esposito}, P.~{Helfenstein}, A.~P.
  {Ingersoll}, R.~{Jaumann}, C.~P. {McKay}, F.~{Nimmo},\ and J.~H. {Waite}
  2009.
\newblock {Enceladus: An Active Cryovolcanic Satellite}.
\newblock In M.~K. {Dougherty}, L.~W. {Esposito}, and S.~M. {Krimigis} (Eds.),
  {\em Saturn from Cassini-Huygens}, pp.\  683.

\bibitem[Steinhilber {\em et~al.}(2012)Steinhilber, Abreu, Beer, Brunner,
  Christl, Fischer, Heikkil{\"a}, Kubik, Mann, McCracken, Miller, Miyahara,
  Oerter, and Wilhelms]{Steinhilber12}
Steinhilber, F., J.~A. Abreu, J.~Beer, I.~Brunner, M.~Christl, H.~Fischer,
  U.~Heikkil{\"a}, P.~W. Kubik, M.~Mann, K.~G. McCracken, H.~Miller,
  H.~Miyahara, H.~Oerter,\ and F.~Wilhelms 2012.
\newblock 9,400 years of cosmic radiation and solar activity from ice cores and
  tree rings.
\newblock {\em Proceedings of the National Academy of Sciences\/}~{\em
  109\/}(16), 5967--5971.

\bibitem[{Stern} {\em et~al.}(2015){Stern}, {Bagenal}, {Ennico}, {Gladstone},
  {Grundy}, {McKinnon}, {Moore}, {Olkin}, {Spencer}, {Weaver}, {Young},
  {Andert}, {Andrews}, {Banks}, {Bauer}, {Bauman}, {Barnouin}, {Bedini},
  {Beisser}, {Beyer}, {Bhaskaran}, {Binzel}, {Birath}, {Bird}, {Bogan},
  {Bowman}, {Bray}, {Brozovic}, {Bryan}, {Buckley}, {Buie}, {Buratti},
  {Bushman}, {Calloway}, {Carcich}, {Cheng}, {Conard}, {Conrad}, {Cook},
  {Cruikshank}, {Custodio}, {Dalle Ore}, {Deboy}, {Dischner}, {Dumont},
  {Earle}, {Elliott}, {Ercol}, {Ernst}, {Finley}, {Flanigan}, {Fountain},
  {Freeze}, {Greathouse}, {Green}, {Guo}, {Hahn}, {Hamilton}, {Hamilton},
  {Hanley}, {Harch}, {Hart}, {Hersman}, {Hill}, {Hill}, {Hinson}, {Holdridge},
  {Horanyi}, {Howard}, {Howett}, {Jackman}, {Jacobson}, {Jennings}, {Kammer},
  {Kang}, {Kaufmann}, {Kollmann}, {Krimigis}, {Kusnierkiewicz}, {Lauer}, {Lee},
  {Lindstrom}, {Linscott}, {Lisse}, {Lunsford}, {Mallder}, {Martin}, {McComas},
  {McNutt}, {Mehoke}, {Mehoke}, {Melin}, {Mutchler}, {Nelson}, {Nimmo},
  {Nunez}, {Ocampo}, {Owen}, {Paetzold}, {Page}, {Parker}, {Parker},
  {Pelletier}, {Peterson}, {Pinkine}, {Piquette}, {Porter}, {Protopapa},
  {Redfern}, {Reitsema}, {Reuter}, {Roberts}, {Robbins}, {Rogers}, {Rose},
  {Runyon}, {Retherford}, {Ryschkewitsch}, {Schenk}, {Schindhelm}, {Sepan},
  {Showalter}, {Singer}, {Soluri}, {Stanbridge}, {Steffl}, {Strobel}, {Stryk},
  {Summers}, {Szalay}, {Tapley}, {Taylor}, {Taylor}, {Throop}, {Tsang},
  {Tyler}, {Umurhan}, {Verbiscer}, {Versteeg}, {Vincent}, {Webbert}, {Weidner},
  {Weigle}, {White}, {Whittenburg}, {Williams}, {Williams}, {Williams},
  {Woods}, {Zangari}, and {Zirnstein}]{Stern15}
{Stern}, S.~A., F.~{Bagenal}, K.~{Ennico}, G.~R. {Gladstone}, W.~M. {Grundy},
  W.~B. {McKinnon}, J.~M. {Moore}, C.~B. {Olkin}, J.~R. {Spencer}, H.~A.
  {Weaver}, L.~A. {Young}, T.~{Andert}, J.~{Andrews}, M.~{Banks}, B.~{Bauer},
  J.~{Bauman}, O.~S. {Barnouin}, P.~{Bedini}, K.~{Beisser}, R.~A. {Beyer},
  S.~{Bhaskaran}, R.~P. {Binzel}, E.~{Birath}, M.~{Bird}, D.~J. {Bogan},
  A.~{Bowman}, V.~J. {Bray}, M.~{Brozovic}, C.~{Bryan}, M.~R. {Buckley}, M.~W.
  {Buie}, B.~J. {Buratti}, S.~S. {Bushman}, A.~{Calloway}, B.~{Carcich}, A.~F.
  {Cheng}, S.~{Conard}, C.~A. {Conrad}, J.~C. {Cook}, D.~P. {Cruikshank}, O.~S.
  {Custodio}, C.~M. {Dalle Ore}, C.~{Deboy}, Z.~J.~B. {Dischner}, P.~{Dumont},
  A.~M. {Earle}, H.~A. {Elliott}, J.~{Ercol}, C.~M. {Ernst}, T.~{Finley}, S.~H.
  {Flanigan}, G.~{Fountain}, M.~J. {Freeze}, T.~{Greathouse}, J.~L. {Green},
  Y.~{Guo}, M.~{Hahn}, D.~P. {Hamilton}, S.~A. {Hamilton}, J.~{Hanley},
  A.~{Harch}, H.~M. {Hart}, C.~B. {Hersman}, A.~{Hill}, M.~E. {Hill}, D.~P.
  {Hinson}, M.~E. {Holdridge}, M.~{Horanyi}, A.~D. {Howard}, C.~J.~A. {Howett},
  C.~{Jackman}, R.~A. {Jacobson}, D.~E. {Jennings}, J.~A. {Kammer}, H.~K.
  {Kang}, D.~E. {Kaufmann}, P.~{Kollmann}, S.~M. {Krimigis},
  D.~{Kusnierkiewicz}, T.~R. {Lauer}, J.~E. {Lee}, K.~L. {Lindstrom}, I.~R.
  {Linscott}, C.~M. {Lisse}, A.~W. {Lunsford}, V.~A. {Mallder}, N.~{Martin},
  D.~J. {McComas}, R.~L. {McNutt}, D.~{Mehoke}, T.~{Mehoke}, E.~D. {Melin},
  M.~{Mutchler}, D.~{Nelson}, F.~{Nimmo}, J.~I. {Nunez}, A.~{Ocampo}, W.~M.
  {Owen}, M.~{Paetzold}, B.~{Page}, A.~H. {Parker}, J.~W. {Parker},
  F.~{Pelletier}, J.~{Peterson}, N.~{Pinkine}, M.~{Piquette}, S.~B. {Porter},
  S.~{Protopapa}, J.~{Redfern}, H.~J. {Reitsema}, D.~C. {Reuter}, J.~H.
  {Roberts}, S.~J. {Robbins}, G.~{Rogers}, D.~{Rose}, K.~{Runyon}, K.~D.
  {Retherford}, M.~G. {Ryschkewitsch}, P.~{Schenk}, E.~{Schindhelm},
  B.~{Sepan}, M.~R. {Showalter}, K.~N. {Singer}, M.~{Soluri}, D.~{Stanbridge},
  A.~J. {Steffl}, D.~F. {Strobel}, T.~{Stryk}, M.~E. {Summers}, J.~R. {Szalay},
  M.~{Tapley}, A.~{Taylor}, H.~{Taylor}, H.~B. {Throop}, C.~C.~C. {Tsang},
  G.~L. {Tyler}, O.~M. {Umurhan}, A.~J. {Verbiscer}, M.~H. {Versteeg},
  M.~{Vincent}, R.~{Webbert}, S.~{Weidner}, G.~E. {Weigle}, O.~L. {White},
  K.~{Whittenburg}, B.~G. {Williams}, K.~{Williams}, S.~{Williams}, W.~W.
  {Woods}, A.~M. {Zangari},\ and E.~{Zirnstein} 2015.
\newblock {The Pluto system: Initial results from its exploration by New
  Horizons}.
\newblock {\em Science\/}~{\bf 350}, aad1815.

\bibitem[{St{\"o}ffler} and {Ryder}(2001){St{\"o}ffler} and {Ryder}]{SR01}
{St{\"o}ffler}, D.,\ and G.~{Ryder} 2001.
\newblock {Stratigraphy and Isotope Ages of Lunar Geologic Units: Chronological
  Standard for the Inner Solar System}.
\newblock {\em \ssr\/}~{\bf 96}, 9--54.

\bibitem[Tazoe {\em et~al.}(2014)Tazoe, Yamagata, Obata, and Nagai]{Tazoe14}
Tazoe, H., T.~Yamagata, H.~Obata,\ and H.~Nagai {2014}.
\newblock {Determination of picomolar beryllium levels in seawater with
  inductively coupled plasma mass spectrometry following silica-gel
  preconcentration}.
\newblock {\em {ANALYTICA CHIMICA ACTA}\/}~{\bf {852}}, {74--81}.

\bibitem[{Wieler} {\em et~al.}(2013){Wieler}, {Beer}, and {Leya}]{Wieler13}
{Wieler}, R., J.~{Beer},\ and I.~{Leya} 2013.
\newblock {The Galactic Cosmic Ray Intensity over the Past 10$^{6}$-10$^{9}$
  Years as Recorded by Cosmogenic Nuclides in Meteorites and Terrestrial
  Samples}.
\newblock {\em \ssr\/}~{\bf 176}, 351--363.

\bibitem[{Zahnle} {\em et~al.}(2003){Zahnle}, {Schenk}, {Levison}, and
  {Dones}]{Zahnle03}
{Zahnle}, K., P.~{Schenk}, H.~{Levison},\ and L.~{Dones} 2003.
\newblock {Cratering rates in the outer Solar System}.
\newblock {\em Icarus\/}~{\bf 163}, 263--289.

\bibitem[{Ziegler} {\em et~al.}(2010){Ziegler}, {Ziegler}, and
  {Biersack}]{SRIM10}
{Ziegler}, J.~F., M.~D. {Ziegler},\ and J.~P. {Biersack} 2010.
\newblock {SRIM - The stopping and range of ions in matter (2010)}.
\newblock {\em Nuclear Instruments and Methods in Physics Research B\/}~{\bf
  268}, 1818--1823.

\end{thebibliography}

\end{document}